\documentclass[english,showpacs,floatfix,12pt]{revtex4}
\usepackage[dvips]{graphicx}
\usepackage{longtable}
\usepackage{amsmath,amssymb}
\usepackage{dcolumn}
\usepackage{latexsym}
\usepackage{babel}
\usepackage[latin1]{inputenc}
\usepackage{color}
\usepackage[colorlinks]{hyperref}
\def\pa{\partial}
\def\al{\alpha}
\def\ga{\gamma}
\def\dl{\delta}
\def\La{\Lambda}

\def\Om{\Omega}
\def\om{\omega}

\def\th{\theta}
\def\sg{\sigma}
\def\vf{\varphi}
\def\vr{\varrho}
\def\wt{\widetilde}

\def\lan{\langle}
\def\ran{\rangle}

\def\ck{\check}

\def\diag{\mbox {diag}}
\def\l{\left}
\def\r{\right}
\def\Up{\Upsilon}

\begin{document}
\title{Nonlinear Spinors as the Candidate of Dark Matter}
\author{Ying-Qiu Gu}
\email{yqgu@fudan.edu.cn} \affiliation{School of Mathematical
Science, Fudan University, Shanghai, 200433, China} 
\date{30th August 2017}

\begin{abstract}
In this paper, we discuss the equation of state for nonlinear spinor
gases in the context of cosmology. The mean energy momentum tensor
is similar to that of the prefect fluid, but an additional function
of state $W$ is introduced to describe the nonlinear potential. The
equation of state $w(a)\lesssim -1$ in the early universe is
calculated, which provides a natural explanation for the negative
pressure of dark matter and dark energy. $W$ may be also the main
origin of the cosmological constant $\Lambda$. So the nonlinear
spinor gases may be a candidate for dark matter and dark energy.
\end{abstract}

\maketitle

\section{Introduction}
\setcounter{equation}{0}

Since the discovery that the expansion of the universe is
accelerating\cite{1}-\cite{weos}, the standard model of cosmology
has shifted from a matter dominated and decelerating expansion
picture to search for the dark matter and dark energy with repulsive
gravity or modification of the general relativity. The present
methods to analyze the observational data include direct measures of
cosmic scales through Type Ia supernova luminosity distances, the
angular distance scales of baryon acoustic oscillation and cosmic
microwave background density perturbations, as well as indirect
probes such as the effect of cosmic expansion on the growth of
matter density fluctuations. The basic building blocks of the
universe is now believed to consist of 4\% baryons, 20\% dark
matter, and 76\% dark energy. The cosmic accelerating expansion
results in a profound mysteries in all of science, with deep
connections to both astrophysics and particle physics.

The simplest choice for such dark energy with negative pressure is
the cosmological constant $\La$, a term consists with Einstein's
equations. This term acts like a fluid with equation of state(EOS)
$p_{\La} =-\rho_{\La}$. Although the $\La$CDM model is basically
consistent with all observational data, the evaluated $\La$ seems
not coincident with the local Newtonian approximation of the
galactic gravity, and a lot of theoretical efforts was paid to
explain its origin and apparent value\cite{lamb1}-\cite{lamb4}.

The widely studied dynamical models to describe dark energy are
mainly the scalar fields, such as quintessence, K-essence, tachyon,
phantom and dilaton,  which are uniformly distributed in the
universe\cite{lamb4}-\cite{dyn}. These models have different
theoretical origin and consequences, and some cases can give
explanation for the observational data and the origin of
$\La$\cite{padm1,padm2,nogo}.

Noting the facts that all fermions are described by spinors and the
uniform scalar field can hardly explain the galactic
structure\cite{gu10}, so the dark spinors may be partially
responsible for dark matter. In this paper we give a detailed
discussion for the EOS of nonlinear spinor gas
 in the context of cosmology. Here the spinors are
quantized and identified by nonlinear potentials, and their state
functions are derived from normal Dirac equations. This work is a
further research of the preceding papers\cite{gu1,gu2,spn0}. Some
similar works were once done in\cite{spn1,spn2,spn3,spn4,spn5,spn6},
where the authors also considered the nonlinear spinor field as
candidate of dark energy, and only one spinor field was taken into
account.

The nonlinear spinors provide us some interesting consequences and
new insights: The negative pressure exists and we have the equation
of state(EOS) $w(z)\sim-1$ in the early universe. The state
functions are only the functions of scale factor but independent of
its derivatives. The calculations show that the nonlinear spinors
can give more natural explanation to the cosmic accelerating
expansion and other observational data.

\section{Equation of state of nonlinear spinor gas}
\setcounter{equation}{0} Define $4\times4$ Hermitian matrices as
follows
\begin{eqnarray}\al^\mu=\left\{\left ( \begin{array}{ll} I & ~0 \\
0 & I \end{array} \right),\left (\begin{array}{ll} 0 & \vec\sg \\
\vec\sg & 0 \end{array}
\right)\right\},~ \ga =\left ( \begin{array}{ll} I & ~0 \\
0 & -I \end{array} \right).\label{e2.11}
\end{eqnarray}
We consider the following Lagrangian for the same kind spinors in
flat spacetime\cite{gu1,gu2}
\begin{eqnarray}
{\cal L}_m =\sum_{n}\l(\phi_n^+(\al^\mu  i\pa_\mu-\mu
\ga)\phi_n+V_n(\ck\ga_n)\r),\qquad \ck\ga_n=\phi_n^+\ga\phi_n
\label{e2.12}\end{eqnarray}  where $\phi_n$ stands for the $n$-th
dark spinor, $\mu>0$ is constant with mass dimension, which takes
one value for the same kind particles. $V_n(\ck\ga_n)$ is the
nonlinear potential term.

The simplest case of dark spinor has the self potential $V_n=\frac 1
2 \om \ck\ga^2_n$, in which $\om>0$ is a constant. The dynamical
equation for $n$-th spinor is given by
\begin{eqnarray}
\hbar i \partial_t\phi_n =\hat H \phi_n,\qquad \hat
H=c[\vec\alpha\cdot\hat p+(\mu c -\om \check\gamma_n
)\gamma].\label{2.1}
\end{eqnarray}
In \cite{gu2,spn0} we computed the static mass $m_n$ and $w_n$ of a
nonlinear spinor as the functions of spectrum parameter $a$(which
cannot be confused with the following scale factor $a$),
\begin{eqnarray}
m_n = \frac{a^2-1}{a^2+1}\mu,\quad w_n\equiv  \frac{1}{2}\om
\int^\infty_0 \check\gamma_n^2\cdot 4\pi r^2 dr=
\frac{a\mu}{a^2+1}\frac{\int^\infty_0 (u^2-v^2)^2\rho^2
d\rho}{\int^\infty_0(u^2+v^2)\rho^2 d\rho}. \label{2.18}
\end{eqnarray}
For the ground state, by Fig.1 in \cite{spn0} we have
\begin{eqnarray}
1<a<2, \quad { E}={(m_n+w_n)c^2} \sim {\mu c^2},\quad m_n\sim
w_n.\label{rhow}
\end{eqnarray}

For dark spinor gas (\ref{e2.12}), in microscopic view, the
classical approximation of energy momentum tensor is given
by\cite{eng}
\begin{eqnarray}
T^{\mu\nu}=\sum_n(m_n u^\mu_n u^\nu_n +w_n g^{\mu\nu})\dl^3(\vec x
-\vec X_n)\sqrt{1-v^2_n},\label{e9}
\end{eqnarray}
where  $u^\mu_n$ is the 4-vector velocity of $\phi_n$, $\vec v_n$
the usual 3-d speed, $\vec X_n(t)$ the central coordinate. In the
case of linear spinor, we have the ideal gas model with
$w_n=0,(\forall n)$, and then we have the usual energy-momentum
tensor for perfect fluid. In case of $w_n>0$, the complete
energy-momentum tensor should be
\begin{eqnarray}
T^{\mu\nu}=(\rho +P)U^\mu U^\nu +( W-P)g^{\mu\nu},\label{e28}
\end{eqnarray}
where the additional term $W$ is a new function of state reflects
the nonlinear potential of particles
\begin{eqnarray}
W= \frac 1 V  \int_V \sum_n w_n \dl^3(\vec x -\vec
X_n)\sqrt{1-v^2_n} d V = \frac 1 V  \sum_{X_n\in V}  w_n
\sqrt{1-v^2_n}, \label{e290}
\end{eqnarray}
which acts like negative pressure. In the above equations we should
distinguish the total energy density $ T^0_{~0}=\rho+W$ with the
mass density $\rho$. They are different concepts for nonlinear
spinor.

In a galaxy the usual pressure $P\sim 0$, by (\ref{rhow}) we have
$\rho\sim W$ and the total EOS in cosmology $w(z)\equiv P_{\rm
tot}/\rho_{\rm tot}\sim-1$ in (\ref{e28})\cite{wamp5,weos}. So the
nonlinear spinor gas can give a natural explanation for the weird
properties of dark matter and dark energy in category of normal
field theory. Since $W$ take the place of $\La$ in Einstein's field
equation, it may be the main origin of the cosmological constant
$\La$. However the function of state $W$ is not a constant in
cosmology, because the relativistic factor $\sqrt{1-v^2_n}$ is
related to temperature $T$\cite{gu8}, and $w_n$ is related to the
scale factor. In what follows we give detailed analysis for these
conclusions.

\section{relations and equations in cosmology}
\setcounter{equation}{0}

In cosmology the scale factor $a(t)$ varies to very large range. In
this case, the potential energy $w_n$ cannot be treated as constant.
It should be the function of $a$. Now we derive $W(a)$ from the
nonlinear Dirac equation coupling with Einstein equation.

In this paper, we adopt the conformal coordinate system for the
universe, then the corresponding Friedmann-Robertson-Walker metric
is given by
\begin{eqnarray}
g_{\mu\nu}=a^2(t)\diag\l[1,-1,-{\cal S}^2(r),-{\cal
S}^2(r)\sin^2\th\r], \label{e2.1}
\end{eqnarray}
where
\begin{eqnarray} {\cal S}=\left \{ \begin{array}{ll}
  \sin r  & {\rm if} \quad  K=1,\\
   r   & {\rm if} \quad  K=0,\\
  \sinh r  & {\rm if} \quad  K=-1.
\end{array} \right. \label{e2.2}
\end{eqnarray}
The  dynamic equation (\ref{2.1}) for each spinor becomes
\begin{eqnarray}
\wt\al^\mu i(\pa_\mu+\Up_\mu)\phi_n=(\mu -V'_n)\ga \phi_n,\quad
\Up_\mu=\left(\frac{3a'}{2a},\frac {{\cal S}'} {\cal S} ,\frac 1 2
\cot\th,0 \right), \label{e2.15}
\end{eqnarray}
where $\wt\al^\mu$ is the coefficient matrix in curved space-time.
It is easy to check the normalization condition holds for each
spinor
\begin{eqnarray}
\int_\Om |\phi_n|^2 a^3 d\Om=1,~~(\forall n), \qquad d\Om={\cal
S}^2\sin\th drd\th d\vf, \label{e2.16}\end{eqnarray} where $d\Om$ is
the comoving volume element independent of $t$. Denote $\hat p^\nu
=i(\pa^\nu+\Up^\nu)$, the energy momentum tensor is given
by\cite{eng}
\begin{eqnarray}
T^{\mu\nu}=\sum_n\left(\frac 1 2 \Re \lan\phi^+_n( \wt\al^\mu \hat
p^\nu+\wt\al^\nu \hat
p^\mu)\phi_n\ran+(V'_n\ck\ga_n-V_n)g^{\mu\nu}\right),
\label{e2.19}\end{eqnarray} where $\Re\lan\ran$ means taking real
part. The classical approximation of (\ref{e2.19}) is just
(\ref{e9}).

In what follows we take $V_n=\frac 1 2 {\om}\ck\ga_n^2$ as example
to derive the equation of state $w(a)$ in usual sense. By
normalizing condition (\ref{e2.16}) we learn that $|\phi_n|^2\to
\dl(\vec x-\vec X_n)$, then in some sense we have $\ck\ga^2_n \sim
\dl^2(\vec x-\vec X_n)$ which depends on the scale of the space when
space deform seriously. To clarify this problem, making scaling
transformation
\begin{eqnarray}
r=\frac {\bar r}{a_0}, \qquad \phi_n= \l(\frac {a_0}{a(t)}\r)^{\frac
3 2}\psi_n,\label{e3.2}
\end{eqnarray} then we get
\begin{eqnarray}
V_n=\frac {\om} 2 \l(\frac {a_0} a\r)^6\wt\ga_n^2,\label{e3.4}
\end{eqnarray}
in which $\wt\ga_n\equiv\psi_n^+\ga\psi_n$. The normalization
condition (\ref{e2.16}) becomes
\begin{eqnarray}
1=\int|\phi|^2 a^3 d \Om =\int|\psi_n|^2 a^3_0 d \Om,\label{e3.6}
\end{eqnarray}
so $\psi_n$ is normalized in the new coordinate system. By
(\ref{e290}), (\ref{e3.4}) and (\ref{e3.6})  we can calculate the
mean value of $W$ as
\begin{eqnarray}
W=  \frac 1 V  \sum_{X_n\in V}\int _\Om V_n a^3 d\Om =  \frac 1
{a^3\Om} \sum_{X_n\in \Om} \l(\frac {a_0}{a}\r)^3 w_n
\sqrt{1-v^2_n}=\frac {\vr\chi}{a^3},\label{e3.10}
\end{eqnarray}
where $\vr = \frac 1 \Om\sum_{X_n\in\Om} m_n$ is the static comoving
mass density of the spinors,
 $\chi=\chi(a)$ is a function defined by
\begin{eqnarray}
\chi=\l(\frac {a_0}{a}\r)^3\frac{\bar w}{\bar m}\l(1-\frac {3\sg
kT}{2(\sg\bar m+kT)}\r),\label{chi}
\end{eqnarray}
$\bar m$ and $\bar w$ are mean mass and potential of all particles
calculated in local Minkowski space-time, $0<\sg<\frac2 3$ is a
parameter determined by energy distribution function. For Maxwell
distribution, we have $\sg=\frac 2 5$.

In \cite{gu8}, we derived the following relations in case $w_n=0$,
\begin{eqnarray}
kT&=&\frac{\sg \bar m }{a}(\sqrt{a^2+b^2}-a),\label{ttoa}\\ \rho &=&
\frac \vr{a^3} \left( 1+\frac {3\sg} {2a}
(\sqrt{a^2+b^2}-a)\right),\quad P=\frac {\vr\sg
b^2}{2a^4\sqrt{a^2+b^2}}.\label{e3.11}\end{eqnarray} where $b$ is a
constant determined by the initial temperature. Substituting
(\ref{ttoa}) into (\ref{chi}), we get
\begin{eqnarray}
\chi=\l(\frac {a_0}{a}\r)^3\frac{\bar w}{\bar m}\l(1 -\frac 3 2
\sg+\frac {3\sg a}{2\sqrt{a^2+b^2} }\r).\label{chi0}
\end{eqnarray}
We have asymptotically  $\chi\to \chi_0 (a_0/a)^3$ if $a\ll a_0$,
where $\chi_0>0$ is a constant.

By (\ref{e28}) we have $T^\mu_{~\nu}=\diag(\rho+W,W-P,W-P,W-P)$.
Substituting (\ref{ttoa}), (\ref{e3.11}) and (\ref{chi0}) into
$T^\mu_{~\nu}$ we get the EOS of nonlinear spinor gas in cosmology
\begin{eqnarray}
w(a)=\frac{P-W}{\rho+W}=\frac{\sg b^2-2\chi
a\sqrt{a^2+b^2}}{\sqrt{a^2+b^2}\l(2\chi a+3\sg\sqrt{a^2+b^2}+2a-3\sg
a\r)}.\label{weos1}
\end{eqnarray}
When $a\ll a_0$ or $\chi\to\infty$ or we get $w\to-1$. This gives a
natural explanation for the negative pressure in the early universe.
(\ref{weos1}) is an increasing function of $a$. This feature is
consistent with the WMAP five-year data\cite{lia}.

\section{discussion and Conclusion}
\setcounter{equation}{0}

According to the above calculation, we get some important
consequences and insights.
\begin{enumerate}
\item By (\ref{e28}), for the nonlinear spinor gas, the nonlinear potential acts as both the negative pressure and positive energy in the energy momentum tensor $T^{\mu\nu}$. When the universe becomes very compact, the term $W$ forms the main part of $T^{\mu\nu}$, and then we have EOS $w(a)\sim-1$. The nonlinear potential may also give an explanation for the origin of cosmological constant $\La$.

\item Although the dynamical equation of the spinors is more complex
than that of the  scalar fields, the state functions of the spinors
are even relatively simpler due to the normalizing condition
(\ref{e3.6}). These functions are only the functions of scale
factor, but independent of its derivatives.

\item Different from the ideal gas and fluid model,
mass-energy density of the spinors is not a monotone decreasing
function of $a$. Spinors have strong responses to the heavily curved
space-time. Some more deeper research should be done.

\item A spinor $\phi_n$ only has a  local micro structure, so one
spinor field can only describe one particle. Some works use one
spinor field to describe global dark energy. This treatment may be
inadequate, and more attention should be paid.

\end{enumerate}

\begin{acknowledgements}
The author is grateful to his supervisors Prof. Ta-Tsien Li  for
encouragement.
\end{acknowledgements}

\end{document}